\documentclass[sigconf]{cidr-2024}

\newtheorem{example}{Example}[section]
\usepackage{enumitem}
\setlist[itemize]{align=parleft,left=0pt..1em}
\usepackage{xspace}
\usepackage{xcolor}
\usepackage{makecell}
\usepackage{hyperref}

\newcommand{\tool}{\textsc{SMARTFEAT}\xspace}
\AtBeginDocument{%
  }

\begin{document}

\title{\tool: Efficient Feature Construction through Feature-Level Foundation Model Interactions}

\author{Yin Lin}
\affiliation{%
 \institution{
  University of Michigan\\
irenelin@umich.edu}
\country{USA}
}
\author{Bolin Ding}
\affiliation{%
 \institution{
  Alibaba Group\\
bolin.ding@alibaba-inc.com}
\country{USA}
}
\author{H. V. Jagadish}
\affiliation{%
 \institution{
  University of Michigan\\
jag@umich.edu}
\country{USA}
}

\author{Jingren Zhou}
\affiliation{%
 \institution{
  Alibaba Group\\
jingren.zhou@alibaba-inc.com}
\country{USA}
}

\renewcommand{\shortauthors}{Yin, et al.}

\begin{abstract}

Before applying data analytics or machine learning to a data set, a vital step is usually the construction of an informative set of features from the data.
In this paper, we present \tool, an efficient automated feature engineering tool to assist data users, even non-experts, in constructing useful features. Leveraging the power of Foundation Models (FMs), our approach enables the creation of new features from the data, based on contextual information and open-world knowledge.
Our method incorporates an intelligent \textit{operator selector} that discerns a subset of operators, effectively avoiding exhaustive combinations of original features, as is typically observed in traditional automated feature engineering tools.
Moreover, we address the limitations of performing data tasks through row-level interactions with FMs, which could lead to significant delays and costs due to excessive API calls.  We introduce a \textit{function generator} that facilitates the acquisition of efficient data transformations, such as dataframe built-in methods or lambda functions, ensuring the applicability of \tool to generate new features for large datasets.
Code repo with prompt details and datasets: (\url{https://github.com/niceIrene/SMARTFEAT}). 
\end{abstract}


\maketitle
\section{Introduction}

Machine learning (ML) plays a crucial role in myriad decision-making processes, ranging from automated policing \cite{richardson2019dirty} to medical diagnosis \cite{kourou2015machine} and pricing plan development\cite{deng2016deep}. Raw data collected through data integration is seldom suitable for direct use for such machine learning (or any other data analytics): there is typically a need for appropriate data wrangling to construct
high-quality features. This process is highly dependent on domain expertise and requires considerable manual effort by data scientists.

\begin{figure}
    \centering
    \includegraphics[width=0.5\textwidth]{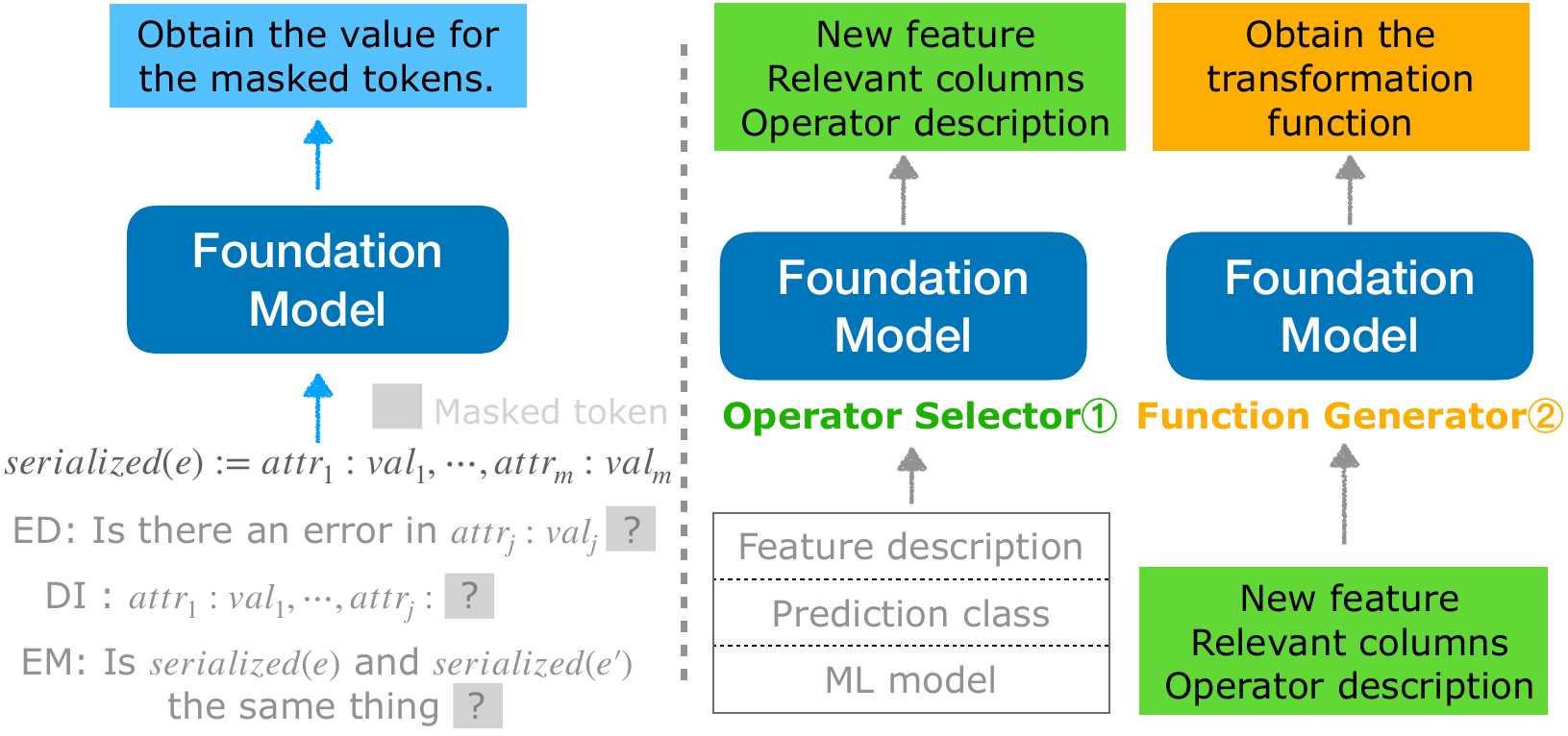}
    \vspace{-2mm}
    \caption{\textnormal{Comparison of row-level interactions to feature-level interactions.}}
    \vspace{-2mm}
    \label{fig:improve}
\end{figure}

To mitigate manual labor, automated approaches have been developed as surveyed in \cite{zoller2021benchmark}. However, traditional automatic feature engineering (AFE) methods typically rely on a predefined set of operators (e.g. scaling a single column, adding two columns), applied directly to the original dataset for generating new features \cite{katz2016explorekit}. Unfortunately, many of the features generated using these methods lack meaningful information and require significant effort for high-quality feature selection \cite{zoller2021benchmark}. 
The limited flexibility in choosing appropriate operators and the inability to capture domain-specific information effectively has led researchers to explore new approaches that can leverage contextual information to enhance feature generation and selection.  Foundation models, as we discuss next, are a natural choice for this purpose.

Recently, the emergence of foundation models (FMs) such as BERT \cite{devlin2018bert}, PaLM \cite{chowdhery2022palm}, and GPT \cite{brown2020language} has brought significant advancements in many applications. 
These models, trained on extensive web-crawled data across diverse tasks, offer the unique capability of adaptation to new tasks without necessitating task-specific fine-tuning \cite{brown2020language}. Consequently, data scientists have been exploring the potential of FMs in handling crucial data-related tasks, including data imputation (DI), error detection (ED), and entity matching (EM), demonstrating SoTA performance \cite{narayan2022can}.

Incorporating FMs, designed to process natural language input and generate corresponding natural language output, into tasks of data manipulation is challenging. The prevailing strategy 
involves serializing and tokenizing each entry in the dataset and using FMs to predict the masked tokens, leveraging their ability to predict the next words as depicted in Figure \ref{fig:improve}. However, conducting row-level completions for large datasets can become impractical due to the time and financial costs associated with excessive interactions with FMs, hindering efficient and scalable applications.

In this paper, we seek to overcome this obstacle and improve the efficiency of integrating FMs into data tasks. Specifically, we explore performing feature-level interactions with FMs, aiming to transform the natural language output of FMs into executable functions that can be applied to construct new features. We demonstrate our idea using the following example.

\vspace{-2mm}
\begin{example} \label{example1}
Consider an insurance company that aims to determine the insurance rate for each customer based on the probability of them filing a new insurance claim within the next 6 months. The dataset in Table \ref{tab:data} includes information such as policyholder demographics, car details, location, and historical insurance records. The prediction class "safe" indicates whether a policyholder is considered safe and less likely to file an insurance claim within the next 6 months.
\end{example}

To enhance prediction performance, 
we present \tool, a practical AFE tool that leverages FMs to predict suitable transformations based on contextual information.  \tool's new features for the example may include:

\noindent
($F1$) \textit{\underline{Bucketized Age}: Groups individuals' ages into predefined bins using a bucketization function.}

\noindent
($F2$) \textit{\underline{Manufacturing Year of the Car}:  Computes the difference between the car's age and the current year.}

\noindent
($F3$) \textit{\underline{Claim Probability per Car Model}: Provides the historical claim probability for each car model by grouping the data accordingly and calculating the average of `Claim in last 6 months'.}

\noindent
($F4$) \textit{\underline{City Population Density}: Extracts population density information from the city feature.}

\noindent
In comparison with traditional AFE tools, \tool offers several compelling advantages:

\begin{table}[tp]
\footnotesize
    \centering
    \caption{\textnormal{Example data set (Prediction class: Safe 1=yes, 0=no).}}
    \vspace{-2mm}
  			\begin{tabular}[b]{|l|l|l|l|l|l|l|}
  			
				\hline
				 \textbf{Sex}&\makecell[c]{\textbf{Age}}& \makecell[c]{\textbf{Age}\\ \textbf{of car}} & \textbf{Make, Model} & \makecell[c]{\textbf{Claim in}\\ \textbf{last 6 month}} & \textbf{City}&\textbf{Safe}
				\\ \hline
				M & 21 & 6 & Honda, Civic& 1 & SF &0\\\hline
			 F & 35 & 2 & Toyota, Corolla & 0 & LA& 1\\\hline
			 M & 42 & 8 & Ford, Mustang & 0 & SEA& 1\\\hline
			 F & 22 & 14 & Chevrolet, Cruze & 1 & SF& 0\\\hline
			 M & 45 & 3 & BMW, X5 & 0 & SEA& 1\\\hline
			 F & 56 & 5 & Volkswagen, Golf & 0 & LA& 1\\\hline
			\end{tabular}
    \label{tab:data}
\vspace{-2mm}
\end{table}

\noindent
\begin{itemize}
\item \textbf{Broader coverage of operators.} 
\tool supports an extensive range of operators, surpassing machine learning-based recommendations \cite{nargesian2017learning}, and pre-defined operators \cite{katz2016explorekit, kanter2015deep}.
The \textit{operator-guided} prompt templates in \tool enable consideration of an extensive set of candidate features. Moreover, leveraging encoded knowledge \cite{narayan2022can} from FMs enables the generation of diverse transformations, such as \texttt{get\_dummies}, \texttt{split}, and user-defined functions. For instance, constructing the bucketization feature ($F1$) with user-defined boundaries can incorporate practical thresholds observed in real-world insurance quotes, like the frequently used threshold of 21 years old.

\item \textbf{Better explainability and efficiency.}  
\tool leverages contextual information within the dataset for feature generation. 
Given descriptions of the current feature set, \tool efficiently interacts with FMs using both \textit{``proposal''} and \textit{``sampling''} strategies \cite{yao2023tree} to generate candidate features. Unlike traditional AFE tools that disregard contextual information and may generate numerous non-meaningful features, \tool selectively considers relevant operators. For example, in the construction of feature ($F2$), \tool exclusively considers the subtraction operator, discarding other binary operators.

\item \textbf{Ability to generate highly correlated features.} 
By providing prompts that specify the prediction class and downstream ML model, \tool tends to generate features that exhibit a notable correlation with the prediction class and are appropriate for downstream ML models. For instance, the creation of the $GroupbyThenAvg$ feature ($F3$) illustrates the claim history of each car model, demonstrating a noteworthy correlation with the "Safe" attribute. Moreover, certain models like k-nearest-neighbors (KNN) tend to perform better when the data is normalized or has similar ranges \cite{peterson2009k}.

\item \textbf{Ability to access external data and resources.}
\tool can leverage external knowledge without explicitly incorporating external sources, as demonstrated in the construction of feature $F4$. Although FMs may have limitations on accessing real-time data, they can provide potential sources and APIs for users to utilize, as further demonstrated in Section \ref{sec:fetching}.
\end{itemize}

In sum, this work makes the following contributions. Firstly, we present an AFE tool that integrates FMs, enhancing the effectiveness of feature engineering. By leveraging contextual information and open-world knowledge, \tool generates a diverse set of relevant features for downstream prediction tasks. Additionally, our approach enables feature-level interactions with FMs and derives values of new features through the generated transformation functions, empowering \tool to handle large datasets.

\section{Background and Related Work}\label{sec:related}

\paragraph{Foundation models for data wrangling.}
Pre-trained language models, also known as foundation models, such as BERT \cite{devlin2018bert}, RoBERTa \cite{liu2019roberta}, GPT-3 \cite{brown2020language}, and ChatGPT, are neural networks trained on large corpora of text data encompassing various tasks. FMs learn the semantics of natural language by predicting the probability of masked words during pre-training and generate text based on the log probability during inference. Typically, FMs consist of billions of parameters and can be applied to a wide range of tasks through fine-tuning or few-shot prompting.
Researchers have recently explored the potential of applying FMs in data management. Narayan et al. \cite{narayan2022can} propose cast data tasks including entity matching (EM), error detection (ED), and data imputation (DI) as prompting tasks to explore FMs' ability to perform classical data wrangling. DITTO \cite{li2020deep} formulates EM as sequence-pair classification, utilizing transformer-based foundation models for enhanced language understanding. RPT \cite{tang2020rpt} investigates pre-trained transformers for data preparation, including EM, DI, and ED. Retclean \cite{ahmad2023retclean} addresses limitations in handling model errors, unseen datasets, and users' private data for ED and DI tasks.
Usually, in the context of these works, FMs are integrated into the data management processes by serializing data entries and predicting masked tokens.

\paragraph{Automated feature engineering.}

A data processing workflow for machine learning or data analytics typically includes critical steps: data acquisition, integration \cite{papadakis2020blocking}, cleaning \cite{abedjan2016detecting}, feature engineering \cite{zoller2021benchmark}, and machine learning training. Our focus lies specifically on the feature engineering step \footnote{We concentrate on the single table scenario, as table joins are typically part of the integration step.}. 
In this context, automated feature engineering (AFE) tools play a vital role in assisting non-experts in constructing high-quality features from raw input data. AFE often employs data transformations to generate new features, thereby enhancing the performance of machine learning predictions. 
Various approaches, such as DSM \cite{kanter2015deep} and OneBM \cite{lam2017one} integrate multiple relations by enumerating potential transformations, while ExploreKit \cite{katz2016explorekit} generates new features using pre-defined operators. 
However, these methods, often referred to as \textit{expansion-selection} methods, tend to create non-meaningful features and require extensive feature selection efforts as the number of generated features is unbounded \cite{khurana2016cognito}. Other approaches, like Cognito \cite{khurana2016cognito} based on tree-like exploration and AutoFEAT \cite{horn2020autofeat} based on iterative search, address some limitations but have low search efficiency. Learning-based methods, such as TransGraph \cite{khurana2018feature} based on Q-learning and LFE \cite{nargesian2017learning} based on MLPs, have also been explored.
FeSTE \cite{harari2022few} proposes using external sources like Wikipedia to enhance classification accuracy but relies on entity matching and may not be applicable to all datasets.
The most relevant work to \tool is CAAFE \cite{hollmann2023llms}, which also leverages FMs to create and validate new feature selections. However, our approach differs from CAAFE in that we define an operator-based search space for generating new features, as opposed to employing an interactive Chain-of-Thoughts \cite{wei2022chain} methodology. 

\section{Proposed Method}
Given a dataset $\mathcal{D}$ consisting of instances $\mathcal{I}$ with original feature set $\mathcal{A} = \{A_1, A_2, \cdots, A_n\}$ and a classification attribute $\mathcal{Y}$, such that $\mathcal{D}_x =\mathcal{I} \times \mathcal{A}$ and $\mathcal{D}_y = \mathcal{Y}$. Our primary objective is to identify a set of suitable operators $Op = \{Op_1, \cdots, Op_m\}$ and their corresponding transformation functions $\mathcal{F} = \{f_1, \cdots, f_m\}$. By applying each transformation function to $\mathcal{D}$, we can obtain a set of candidate new features $\{A_1^{cand}, \cdots, A_m^{cand}\}$ for all $f_i \in \mathcal{F}$. 

\subsection{Overview} \label{sec:overview}
The feature generation process of \tool is similar to other AFE tools \cite{katz2016explorekit, khurana2016cognito, kanter2015deep} that utilize a search process to iteratively create new features and incrementally enhance the existing feature set. These newly generated features are subsequently taken into consideration for the ongoing generation of additional features.

Figure \ref{fig:improve} illustrates, in each iteration, the process by which \tool searches for and generates each new feature, employing two core components: the \textit{operator selector} and the \textit{function generator}, both supported by foundation models (FMs).

The input to \tool comprises three elements: \textit{(1) dataset feature description, (2) prediction class, and (3) downstream classification model.} 
The \textit{dataset feature description} typically contains a belief description of the feature content, data type, and potentially the data domain of categorical features. This information can often be found in the data card of each open-source dataset, such as those available on platforms like Kaggle. We assume that, in most cases, the feature names are descriptive and this information is accessible or can be user-generated. We will also explore \tool's performance with minimal input, consisting only of the feature names in Section \ref{sec:exp}.

The \textit{prediction class} indicates the target variable that the downstream model aims to predict. Lastly, we specify the \textit{classification model} used downstream. While our discussion mainly focuses on the downstream task of binary classification, our model can adapt to constructing features for other downstream applications with minor adjustments to the prompts.

The \textit{operator selector} (\textcircled{1}) chooses suitable operators and generates the descriptions for new features.
It encodes a set of prompt templates corresponding to the operators considered for generating new features.  Given the input, the operator selector uses the prompt for the current operator to interact with the FM model. It selects the appropriate operators and provides the (i) {\it name of the new feature}, the (ii) {\it relevant columns} to compute the new feature, and the (iii) {\it new feature description}. These three outputs serve as input for the function generator to generate the transformation function.

The function generator (\textcircled{2}) seeks to obtain the optimal transformation function or provide necessary information, such as data sources, by interacting with an FM model.  \tool utilizes the state-of-the-art FM interaction toolkit, LangChain\footnote{\url{https://www.langchain.com/}}, to parse the output and then automatically applies the transformation function to the dataset. Once the new feature is successfully generated, both the feature name and its description (generated by the operator selector) are included in the {\it dataset feature description} and used to generate additional features in the next iteration.

\begin{figure}[tp]
    \centering
    \includegraphics[scale= 0.54]{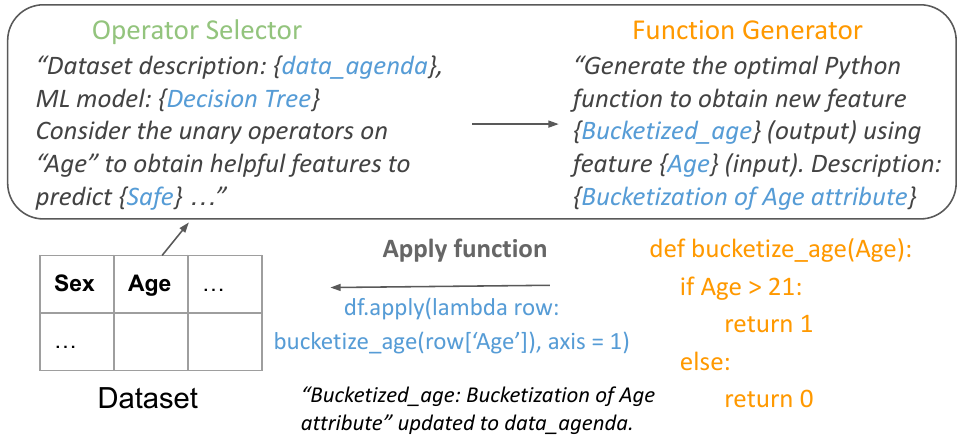}
    \vspace{-4mm}
    \caption{\textnormal{Illustrative example: constructing {\it Bucketized Age}.}}
    \label{fig:descritizer}
    \vspace{-2mm}
\end{figure}

Consider the example in Figure \ref{fig:descritizer}. The current operator being explored by the operator selector is a unary operator. Interacting with its FM model, the operator selector generates the output, including a feature name (\texttt{Bucketized\_age}), a feature description, and the relevant column(s). The function generator then uses this output to obtain an executable function, applies it to the original dataset, and updates the data agenda.
We next delve into the details of the two components in Section \ref{sec:operator} and Section \ref{sec:fetching}, respectively.

\subsection{Operator-guided feature generation} \label{sec:operator}

The goal of this component is to efficiently generate candidate features without exhaustively enumerating all potential combinations.
In this subsection, we first discuss a set of operators used to guide the generation of new features. Then, we explain how we prompt the FM and generate the candidate feature set.

\begin{table*}[t]
    \centering
    \caption{\textnormal{Example of prompt templates and outputs for operator selector.}}
    \label{tab:op_selection}
    \vspace{-2mm}
    \begin{tabular}{cccc}
        \toprule
        \textbf{Strategy} & \textbf{Operator} & \textbf{Prompt template} & \textbf{FM output} \\
        \midrule
        \makecell{Proposal} & \makecell{Unary} & \makecell[l]{{\it `` $\cdots$ Consider the unary operators on the attribute \{org\_attr\} that can }\\ {\it generate helpful features to predict \{y\_attr\}. }\\{\it List all possible appropriate operators, and your confidence levels}\\ {\it (certain/high/medium/low) $\cdots$''}} & \makecell[l]{$Op_1$ ((certain/high/...): desc. 1  \\$Op_2$ ((certain/high/...):  desc. 2 \\ $\cdots$ } \\
        \hline
        \makecell{Sampling} & \makecell{High-order} & \makecell[l]{{\it `` $\cdots$ Generate a groupby feature for predicting \{y\_attr\} by applying }\\ {\it `df.groupby(groupby\_col)[agg\_col].transform(function)'.}\\
{\it Specify the groupby\_col, agg\_col, and the aggregation function.''}} & \makecell[l]{\{groupby\_col: [cols] , agg\_col: col, \\function: mean/max/ $\cdots$\}} \\        
        \bottomrule
    \end{tabular}
\end{table*}

\paragraph{Operator types}
We consider four types of operators for generating candidate features: unary, binary, high-order, and extractor. For each operator, we use a prompt template to interact with the FM.

\textbf{Unary operators} encompass normalization, bucketization, and a set of unary operations such as getting dummies and date splitting. The operator selector does not determine the specific transformation function to be selected, for example, whether to use min-max scaling or standardization for normalization, or what bucket boundaries to set for the bucketization. The focus is solely on assessing the types of unary operations that are beneficial, leaving the function selection to the second phase.

\textbf{Binary operators} include four basic arithmetic operations: $+, -, \times, \div$. We leave the more complex combinations of two original features to extractor operators.

\textbf{High-order operators}, where we consider the $GroupbyThenAgg$ operation. We consider original features that are capable of categorizing data into distinct subsets as valid candidates for $Groupby$ columns and features containing numerical information that can be aggregated as valid aggregate columns. The aggregate functions include mean, max, average, and others, and we allow the FM to choose the optimal function.

\textbf{Extractors} extract information that cannot be obtained from the previous operators. They handle more complex transformations and feature extractions, such as computing an index as a weighted combination of several features.  Additionally, they can leverage library functions and assist in extracting information from external sources, such as obtaining the population density for each city in the motivating example.

\paragraph{Prompting FM for operator selection.}

As discussed in Section \ref{sec:overview}, the operator selector contains a set of prompt templates for each type of operator. 
Based on the characteristics of different operators, we employ two prompting strategies in our search for new features: the \textit{proposal strategy} and the \textit{sampling strategy} \cite{yao2023tree}, which are encoded into the prompt templates of the operators. 

In the proposal strategy, the FM is prompted to propose all potential candidates for $Op$. The candidates are drawn from FM's output by $[Op^{(1)}, \cdots, Op^{(i)}] \sim p^{\text{propose}}(Op^{(1, \cdots, i)}|z_{\text{descr}}, z_{\text{y}},z_{\text{model}})$. From these proposals $Op^{(1, \cdots, i)}$, \tool selects the most probable options, generating descriptive details, relevant columns, and a feature name. This strategy is more effective when dealing with relatively smaller search spaces because the time for searching is limited, and it also avoids duplication. For instance, when exploring the unary operators, we can apply the proposal prompting strategy to each original feature to propose potential unary operators that can be applied to the feature.

Table \ref{tab:op_selection} presents the prompt template and the FM output for the unary operator. Based on the FM output, \tool then selects the operators with {\it certain} or {\it high} confidence to generate new features. \tool parses the output to obtain the new feature name as \texttt{``OpName\_OrgAttr''}, the feature description as the operator description, and the relevant columns as the \texttt{[OrgAttr]}.

In the sampling strategy, the FM is prompted to provide one candidate operator at a time, i.i.d. sampled from a Chain of Thought, i.e.,  $Op^{(i)} \sim p^{\text{sample}}(Op| (z_{\text{descr}}, z_{\text{y}}, z_{\text{model}})$. The sampling method works better when the generation space is rich. For instance, for high-order operators, the selection of $Groupby$ columns can grow exponentially with the number of categorical features. In such cases, the sampling strategy leads to a more efficient generation and a more diverse set of candidates.

In \tool, users can set a sampling budget for feature generation and a threshold for generation errors. The sampling process continues until the budget for sampling or the threshold for generation errors (invalid/repeated features) is reached.

Table \ref{tab:op_selection} presents the prompt template and the FM output for the high-order operator. The FM returns an operator with the selected $Groupby$ columns, the aggregate column, and the aggregate function. Subsequently, \tool parses the output and returns the transformation function as the feature description and the feature name as \texttt{``GroupBy\_Gcol\_func\_Acol''}. The $Groupby$ columns and the aggregate column are included as relevant columns.

\paragraph{Generating the candidate feature set.}

We discuss how operator-guided feature generation works, which aims to maximize the coverage and efficiency of the generation process. We begin by exploring unary operators for each original feature using the proposal strategy. Based on the original and unary features, we apply binary and high-order operators using the sampling strategy. 
Lastly, we consider extractors that can operate on multiple inputs using the sampling strategy, further enriching the current feature set.

We note that prediction performance improvement can also benefit from removing features. In \tool, we employ a heuristic for dropping features: if an original feature undergoes a unary transformation and is not used by any other operators, we consider the original feature less important and, therefore, remove it from the feature set. The exploration of utilizing FMs for feature removal is left as future work.

\subsection{Transformation function generation}\label{sec:fetching}
After identifying candidate operators, the next step involves generating the transformation functions that compute the values of the new features.
For each candidate, the function generator initiates an interaction with the FM to decide whether a transformation function can be derived, leading to three possible scenarios.

Firstly, if a transformation function can be derived, \tool generates it using the relevant columns as input and the new feature as the output. In most cases, interaction with FM is needed to obtain the most appropriate and efficient transformation function. 
For instance, employing FMs can assist in selecting suitable buckets for bucketization and importing necessary library functions as required.
In the case of the high-order operator, the function generator can construct the transformation function directly from the output of the operator selector without the need to interact with the FM.

In certain situations, explicit functions may not be obtainable, requiring row-level text completion to calculate the new feature value, such as extracting the approximate population density for each city. To address this, we serialize the data entries and append the new feature name and a masked token as $\underline{A_1: v_1, \cdots, A_k:v_k, A_{\text{new}}: ?}$, and then query the FM to generate the new feature values. However, for large datasets, the cost of obtaining these values through API calls could be prohibitive. Hence, we provide users with several examples and allow them to decide if the new features are valuable enough considering the associated cost.

Lastly, in situations where the function generator cannot produce a suitable transformation function or text completion is not applicable, the function generator 
suggests potential data sources to assist data users in constructing the new feature.

\paragraph{Evaluating generated features.}
We implemented a basic verification mechanism to ensure the quality of the features derived from FM-generated code. After obtaining the feature values, we perform feature selection to remove features that are highly null, single-valued, or dummy variables derived from high-cardinality original features. This feature selection process enhances the reliability and effectiveness of the generated features.

\section{Evaluation} \label{sec:exp}

\begin{table}[]
\footnotesize
    \centering
    \caption{\textnormal{Dataset statistics.}} \label{tbl:data}
    \vspace{-3mm}
  			\begin{tabular}[b]{|l|l|l|l|l|}
				\hline
		 &\# of cat. attr & \# of num.  attr	& \# of rows & field
            \\ \hline				     
            Diabetes & 0 & 9& 769& Health
            \\ \hline
            Heart & 7 & 7& 3,657& Health
            \\ \hline
            Bank &  8& 10& 41,189 & Finance
            \\ \hline
            Adult & 8 & 6&  30,163& Society
            \\ \hline
             Housing  & 1 & 8& 20,641& Society
            \\ \hline
            Lawschool& 5 & 7& 4,591 & Education
            \\ \hline
            West Nile Virus &  3& 8& 10,507 & Disease
            \\ \hline
            Tennis & 0 & 12 &  944& Sports
		\\ \hline	     
			\end{tabular}
    \label{tab:statistics}
\vspace{-4mm}
\end{table}

We explore the performance of various downstream classification models when using features generated by \tool with datasets from diverse fields, including health, finance, society, education, disease, and sports. We compare these against the state-of-the-art AFE methods.

\begin{table*}[tp]
\small
\centering
     \caption{\textnormal{Comparison of the average AUC values ($\uparrow$) of different ML models between \tool and baseline methods.}} \label{fig:avg}
     \vspace{-4mm}
\begin{tabular}{lcccccccc}
\toprule
 \textbf{Methods}& Diabetes & Heart & Bank & Adult & Housing & Lawschool & West Nile Virus & Tennis \\
\midrule
Initial AUC &82.20 &	67.38 &	91.46 & 76.81 &	86.72&	\textbf{84.00} &	78.96 &	77.93\\
\tool &\textbf{86.76 (+4.3\%)} &	\textbf{72.15 (+7.0\%)} &	91.47 ($\approx$) &\textbf{87.00 (+13.3\%)}	& \textbf{92.19 (+6.3\%)} &	83.68 (-0.4\%) &	\textbf{82.12 (+4.0\%)} &	87.39 (+9.5\%)\\
CAAFE &   -	&  69.67 (+3.4\%) &	\textbf{\underline{91.73 (+0.3\%)}} &	\underline{83.10 (+8.2\%)} &	\underline{92.15 (+6.3\%)} & 83.86 (-0.2\%)&	80.11 (+1.8\%) &	\textbf{88.50 (+13.6\%)}\\
Featuretools & 82.24 ($\approx$)	& 66.78 (-0.9\%) &	91.04 (-0.5\%) &	73.85 (-3.9\%) &	79.47 (-8.1\%)&	83.82 (-0.2\%) &	73.12 (-7.4\%) &	81.29 (+4.3\%) \\
AutoFeat & 75.24 (-8.4\%)& 64.92 (-3.7\%)&  - & - & 77.63 (-10.5\%)& - & 70.90 (-10.2\%)& 71.73 (-8.0\%)\\
\bottomrule
\end{tabular}
\vspace{-2mm}
\end{table*}

\begin{table*}[tp]
\small
\centering
     \caption{\textnormal{Comparison of the median AUC values ($\uparrow$) of different ML models between \tool and baseline methods.}} \label{fig:median}
     \vspace{-4mm}
\begin{tabular}{lcccccccc}
\toprule
 \textbf{Methods}& Diabetes & Heart & Bank & Adult & Housing & Lawschool & West Nile Virus & Tennis \\
\midrule
Initial AUC &83.18 &	69.19 &	92.77 &	80.63 &	91.28 & 83.73 &	77.66 &	80.41 \\

\tool & \textbf{87.78 (+5.5\%)} &	\textbf{71.70 +(3.6\%)} &	92.86 ($\approx$)& 86.97(+7.9\%) &	90.97 (-0.3\%)&	83.32 (-0.5\%) &	\textbf{82.06 (+5.7\%)} & 88.06 (+9.5\%)\\
CAAFE & -	& 70.87 (+2.4\%) &	\textbf{\underline{93.06 +(0.3\%)}} &	\textbf{\underline{87.00 (+7.9\%)}} &	\textbf{\underline{92.84 +(1.7\%)}} &	\textbf{83.77 ($\approx$)}&	80.90 (+4.2\%) &	\textbf{89.51 (+10.9\%)}\\
Featuretools  & 82.78 (-0.5\%)&	69.37(-0.3\%) &	91.06 (-1.8\%) &	68.91 (-14.5\%)&	73.39(-19.0\%) &	83.74($\approx$) &	75.71 (-2.5\%) &83.03 (+3.3\%)\\
AutoFeat & 84.20 (+0.1\%)& 70.42 (+1.7\%)&  - & - & 75.65(-17.1\%)& - & 76.53 (-1.4\%)& 67.83 (-15.6\%)\\
\bottomrule
\end{tabular}
\vspace{-2mm}
\end{table*}

\subsection{Experimental setup}
\paragraph{Datasets.} We conducted the evaluation on eight supervised binary classification datasets, publicly available on Kaggle\footnote{\url{https://www.kaggle.com/competitions/}}. The summary of dataset statistics is shown in Table \ref{tbl:data}. 
We assessed the effectiveness of the FM in effectively handling column contexts across various domains. 
We randomly partitioned each dataset into 75\% for training and 25\% for testing and used 10-fold cross-validation. Prior to conducting the feature engineering process, we executed standard data cleaning procedures on the datasets, including the removal of missing values (dropna) and the factorization of categorical features.

\paragraph{Metrics.} To assess the effectiveness of new features, we employed five downstream machine learning classification algorithms from sklearn\footnote{\url{https://scikit-learn.org/stable/supervised_learning.html}}, which include Linear Regression (LR), GaussianNB (NB), Random Forest (RF), and Extra Tree (ET). Additionally, we incorporated a deep neural network (DNN) to demonstrate the potential improvement brought about by the constructed features, even though DNNs inherently possess the capability to learn deep features \cite{lecun2015deep}. For all models, we utilized default parameter settings. The neural network architecture comprised two hidden layers, each consisting of 100 units and employing the ReLU activation function. In our evaluation, we considered the Area Under the ROC Curve (AUC) as the primary performance metric. 

\paragraph{Baselines.} We compared \tool with the SoTA AFE tools as discussed in Section \ref{sec:related}.  
However, only DSM \cite{kanter2015deep},  AutoFEAT \cite{horn2020autofeat}, and CAAFE \cite{hollmann2023llms} offer publicly accessible implementations. Consequently, our experiments for comparison are based on these works. 
For DSM, we utilized its community-supported tool called \textbf{Featuretools} \footnote{\url{https://featuretools.alteryx.com/en/stable/}}, which exhaustively generates new features using pre-defined operators and incorporates feature selection to eliminate highly correlated, highly null, and single-value features.
AutoFEAT (referred to as \textbf{AutoFEAT} \footnote{\url{https://github.com/cod3licious/autofeat/tree/master}}) adopts a different approach by constructing a large set of non-linear features and subsequently performing a search algorithm to select an effective subset.
\textbf{CAAFE}\footnote{\url{https://github.com/automl/CAAFE}} is an AFE tool that utilizes FMs to generate Python code for data transformations.  It includes a validation step: newly generated transformations are retained only if they demonstrate performance improvement on the validation set.
For the implementation of Featuretools, we specifically utilized the primitives "add\_numeric," "multiply\_numeric," and "agg\_primitive," while retaining default settings for other parameters. For AutoFeat, we used all default parameters. For CAAFE, we used its implementation with GPT-4 and 10 feature generation iterations as in \cite{hollmann2023llms}.

In \tool, we leveraged OpenAI's GPT-4 as the FM in the operator selector. For the function generator, we used the default OpenAI model in LangChain (GPT-3.5-turbo) due to its comparable performance and better efficiency. We employed the zero-shot prompting method in all FMs. The sampling budget for operators using the sampling prompting strategy was set to 10.

All experiments were conducted on a macOS system with a 1.4 GHz Quad-Core Intel Core i5 processor and 16GB of memory. We established a time limit of one hour for all experiments.

\subsection{Evaluation results}
\paragraph{Classification result.}
Table \ref{fig:avg} and \ref{fig:median} present the results of the evaluation in terms of AUC values. The AUC improvement in percentage compared with the initial AUC is shown in parentheses. We highlighted the best-performing approaches in bold and underlined the baselines that do not support all ML models.

We reported the average and median AUC values across all classification models. The results indicate that the FM-assisted methods, \tool, and CAAFE consistently achieve better performance compared with other baselines.
\tool achieves a performance improvement of the average AUC for up to 13.3\%, outperforming the other baselines in 5 out of 8 datasets (Table \ref{fig:avg}). However, we observed that the performance improvement on the \textit{Bank} dataset and the \textit{Lawschool} dataset is not evident, as also observed in the other baselines. This is because, in these two datasets, the original features are well-constructed, making feature engineering less impactful for performance enhancement.

CAAFE improves the performance of the initial AUC on all datasets. This is because it uses the validation set to evaluate the effect of each newly generated transformation and only retains the ones that improve the model performance, effectively preserving the helpful transformations. However, this step would be impractical when working with large datasets. CAAFE experienced a timeout on the DNN model on three large datasets, \textit{Bank}, \textit{Adult}, and \textit{Bank}. CAAFE outperforms \tool mainly on datasets that consist of more numerical features, such as the \textit{Tennis} dataset. This is because, without the operator selector, we observed that the proposed transformations mainly perform combinations of numerical attributes.
In addition, CAAFE samples the feature values, which also helps enhance the effectiveness of the transformations. However, for datasets where diverse types of new features would be beneficial, such as the \textit{West Nile Virus} dataset, \tool can generate more helpful features. CAAFE failed on the \textit{Diabetes} dataset in that it suggested divide-by-zero transformations without handling the NAN values and caused the ML models to fail.

On the other hand, the transformations performed by Featuretools and AutoFeat are agnostic to the dataset context and the prediction task. In some cases, this results in the new features being less suitable for improving prediction performance, sometimes leading to a decrease in the AUC values for these baselines.

\vspace{-2mm}
\paragraph{Efficiency.}
Feature generation is a task performed in concordance with human designers. As such, we need systems to take no more than a few minutes. 
On all datasets, both \tool and Featuretools finished in well under 10 minutes.  This was not the case for Autofeat, as it did not finish within our 60-minute time-out for the \textit{Bank} and \textit{Adult} datasets. As discussed earlier, CAAFE also experienced a time-out on large datasets and complex models due to the validation step.  It also incurs a relatively higher time compared to \tool and Featuretools in general.

\vspace{-2mm}
\paragraph{Feature importance.}
We now delve into the features generated by different approaches to assess their usefulness. We focus on the \textit{Tennis} dataset where most of the approaches demonstrate effectiveness. We assessed the feature importance of all features, including both the new features and the original feature set.
We evaluated three feature selection metrics provided by sklearn: the Information Gain (also known as mutual information), Recursive Feature Elimination, and Feature Importance metric based on tree-based selection (indicator for Gini index). We excluded statistics like $\chi^2-$test or F-value as they are only applicable to either categorical or numerical attributes. Specifically, we examined the percentage of new features in the top-10 important features identified by each metric. For example, if \tool has IG@10 = 80\%, it means that 8 out of the top 10 ranked features under the information gain metric are new features generated by \tool. A higher percentage indicates the generation of more useful features.

As shown in Table \ref{tbl:top10}, CAAFE generates a small number of new features, mainly due to the removal of features in the validation step. However, all new features generated by CAAFE are considered important under all metrics.
In comparison with the other two baselines, \tool generates a smaller number of new features. This is attributed to the operator selector, which intelligently selects a subset of operators based on column context, avoiding an exhaustive enumeration of all possible features. \tool demonstrates effectiveness in obtaining helpful features.
Featuretools also successfully obtains important features but with a relatively large number of generated features. AutoFeat initially acquires a vast set of new features, but the feature selection steps only retain 5 features, which are shown less helpful under these metrics.

\vspace{-2mm}
\paragraph{Ablation study.}
We conducted an ablation study to show how different operators in \tool contribute to improving prediction performance. Using the \textit{Tennis} dataset as an example, we assessed the AUC values obtained by adding the new features generated by each type of operator.

As shown in Table \ref{tab:ablation}, the binary and extractor operators bring performance improvement in most cases, particularly for NB, RF, and ET. In this dataset, the features introduced by the extractor are mainly index-like attributes computed from the combination of a set of attributes. LR's performance remains almost unaffected, indicating that feature engineering does not provide substantial benefits in this case. For DNN, we observed that almost all types of features contribute to enhancing the prediction results. This is because \textit{Tennis} is a relatively small dataset, where simple models with well-constructed features usually perform better.

However, it's important to note that this trend may not be consistent across other datasets. For example, in the \textit{West Nile Virus} dataset, the most beneficial features are those generated by the high-order operators.

\paragraph{Impact of Feature Descriptions.} We also investigate the significance of having informative feature descriptions as input. To assess this, we conducted a comparative experiment on the \textit{Tennis} dataset using only feature names without descriptions in \tool. The feature names in the \text{Tennis} dataset are less descriptive, featuring abbreviations like \texttt{FSW.1} to represent "First Serve Percentage for player 1." In this experiment, the AUC score dropped to 77.86 (-1.4\%) for the average and 79.39 (+2.2\%) for the median. The experimental results emphasize the importance of including meaningful feature descriptions, particularly when clear and informative feature names are not available.

\begin{table}[tp]
\footnotesize
\caption{\textnormal{Percentage of top-10 important features ($\uparrow$) generated by different methods under varying feature selection metrics, \textit{Tennis}.}} \label{tbl:top10}
\vspace{-2mm}
\centering
\begin{tabular}{lcccc}
\toprule
 & \tool &CAAFE & Featuretools & AutoFeat  \\
 \hline
 \textbf{\# generated features} & 25 & 5 &  89 (sel-35)& 1978 (sel-5) \\
\midrule
\textbf{IG@10} & 90\% & 50\% (all) &90\% & 10\% \\
\textbf{RFE@10} & 80\% &50\% (all)&90\% &30\% \\
\textbf{FI@10} & 80\%& 50\% (all) &90\% & 30\%\\
\bottomrule
\end{tabular}
\vspace{-2mm}
\end{table}

\begin{table}[tp]
\footnotesize
\centering
\caption{\textnormal{Ablation study on operators in \tool across different downstream ML models, \textit{Tennis}.}}
\label{tab:ablation}
\vspace{-4mm}
\begin{tabular}{lllllll}
\toprule
\text{}& \text{Initial}& \text{+Unary} & \text{+Binary} & \text{+High-order} & \text{+Extractor} & \text{all} \\ 
\midrule
LR &        88.17	&88.27&	88.51	&88.22 	&88.53	&88.06 \\ 
NB  &  66.85&	65.16	&79.68	&66.49	&90.00	&84.05\\ 
RF   &   80.41&	81.17&	87.38&	80.15 &	89.88 &	89.56   \\
ET   &     79.14 & 	75.14 & 	88.02 & 77.56	&90.04 &88.86     \\ 
DNN   &     84.50 &	87.31 &87.57&	86.08 &	86.92	& 86.46 \\ 

\midrule
\textbf{Avg}   &  79.81& 79.41 &86.31	& 79.70 &	89.07 &	87.39    \\ 

\bottomrule
\end{tabular}
\vspace{-2mm}
\end{table}

\section{Final Remarks}
Foundation models hold the promise of leveraging contextual information and open-world knowledge for many data management and analysis tasks, such as feature generation. 
However, it is infeasible to feed these models with simple linearizations of relational tables for large databases, for reasons of efficiency and cost, even if the FMs were able to accept such large inputs. In this paper, we proposed operator-guided feature generation, coupled with feature-level interactions with FMs, to achieve both high efficiency and comprehensive coverage. 

Moreover, FMs are susceptible to unpredicted errors, arising from limited access to data context or the inherent generative nature of FMs. To mitigate these errors, we employ feature selection to reduce the risk of generating low-quality features. While our experiments demonstrate the advantages of feature generation, further improvements in error correction and detection are areas for future research.
Integrating Foundation Models into data systems also poses challenges, especially when FMs rely on natural language input. Our paper primarily focuses on feature generation tasks, exploring one potential solution by discovering potential operators and their corresponding transformations. However, for more complex data-wrangling tasks that involve exploring data content and enabling feature-level interactions, discussions on obtaining informative data descriptions become essential.

\vspace{-2mm}
\section*{Acknowledgements}
This research was supported in part by NSF under grants 1934565, 2106176, and 2312931.

\bibliographystyle{abbrv}
\bibliography{acmart}


\end{document}